\documentclass[aps,twocolumn,showpacs]{revtex4}%
\usepackage{amsfonts}
\usepackage{amsmath}
\usepackage{amssymb}
\usepackage{graphicx}
\usepackage{color}%
\providecommand{\U}[1]{\protect\rule{.1in}{.1in}}
\begin{document}
\title{All-optical photon transport switching in a passive-active optomechanical system}

\author{Lei Du$^{1}$, Yi-Mou Liu$^{1}$, Yan Zhang$^{1,\ast}$, and Jin-Hui Wu$^{1,\dagger}$}

\affiliation{$^{1}$Center for Quantum Sciences and School of
Physics, Northeast Normal University, Changchun 130117, P. R. China}
\affiliation{$^{\ast}$zhangy345@nenu.edu.cn}
\affiliation{$^{\dagger}$jhwu@nenu.edu.cn}
\date{\today }


\begin{abstract}
We propose a feasible scheme to realize all-optical photon transport switching in a passive-active optomechanical system, consisting of
one ordinary passive cavity, one active cavity and one common movable membrane oscillator of perfect reflection, driven by two strong control fields
and two weak probe fields symmetrically. By means of the gain effect of the active cavity, many novel and valuable phenomena arise, such as
frequency-independent perfect reflection (FIPR) first proposed in this paper, adjustable photon bidirectional transport, phase-dependent
non-reciprocity and so on. The relevant parameters used for controlling the all-optical switching are precisely tunable by adjusting the strengths
of control fields and the relative phase of probe fields. Furthermore, tunable fast and slow light can be realized in our system by accurately
adjusting relevant parameters which is readily and feasible in experiments. These novel phenomena originate from the effective optomechanical coupling
and the gain effect provide a promising platform for photonic devices, quantum network node fabrication and quantum information process (QIP).
\end{abstract}
\maketitle

\section{Introduction}

Optomechanical systems attracted intense interest and numerous investigations in the past few
decades due to the superiority that plenty of quantum phenomena can be exhibited in such macroscopic
devices, which provides a promising platform to explore quantum science and the correlation
between classical and quantum issues \cite{optooverview,quantumopto}. As a interdiscipline of
microphysics and quantum optics, which primarily studying the interactions between the electromagnetic
fields and mechanical systems, it has developed rapidly that series of advances born in recent years,
e.g., optomechanically induced transparency (OMIT) \cite{OMITscience,OMITnature,OMITpra2013,doubleOMIT2014},
multiple quantum entanglement \cite{YDWangprl2013,tripartite2017}, signal absorption and amplification
\cite{ab-am1,ab-am2,ab-am3,ab-am4}, photon blocked \cite{pblocked1,pblocked2,pblocked3}, optomechanically 
induced non-reciprocity \cite{reci1,reci2,reci3} and so on, which laying a solid foundation for controllable 
photon transport and quantum information process via macroscopical optomechanical devices.

Recently, quite a lot of creative and satisfactory phenomena arise in the systems containing active (gain) 
subparts, especially in so-called parity-time ($\mathcal{PT}$)-symmetric systems, exhibiting extraordinary 
advantages compared with those in conventional passive-passive systems \cite{beamPT,extend1,extend2,extend3}. 
The successful realization of $\mathcal{PT}$ symmetry in optical systems, including optomechanical systems, 
leads to so many novel breakthroughs recently, e.g., mechanical $\mathcal{PT}$ symmetry \cite{ptoslwl,ptosxxw}, 
$\mathcal{PT}$-symmetry-breaking induced ultralow threshold chaos \cite{ptchaos}, photon laser \cite{ptplaserjh,ptlaserhb}, 
optomechanically induced transparency (OMIT) \cite{ptOMITlwl,ptOMITjh} and enhanced photon blocked effect \cite{ptpblocked}.
In a recent study \cite{lylpra2017}, Liu \emph{et.al.} proposed a single-cavity optomechanical system with
an active mechanical resonator, by modifying the gain and loss of the mechanical resonator, a tunable optical
signal amplification and absorption, untraslow light with a more robust group delay can be realized. In view 
of all above, it seems that a passive-active system can induce more beneficial properties compared with 
conventional passive-passive systems in some respects.

In this paper, we study a three-mode passive-active optomechanical system includes two indirectly coupled
cavities and a common membrane oscillator which couples to the two cavity modes via optomechanical coupling
due to radiation pressure. Our passive-active system exhibits a controllable photon transport by adjusting the
strengths of control fields, relative phase of probe fields and the gain of the active cavity where all the
relevant parameters are readily and precisely changed in experiments. It is worth pointing out that the frequency-independent
perfect reflection (FIPR), which arises in the gain-loss-balanced case, is very different from the phenomena
in passive-passive systems such as coherent perfect absorption (CPA), coherent perfect transmission (CPT) and
so on which arise only in specific parameter regions and especially rely on the frequency of probe field \cite{Huang,XBYan},
is independent of all other parameters once the specific condition is met. Moreover, one can further realize tunable
fast and slow light in each case discussed in this paper, which is effortless and feasible in experiments.

\section{Model and Methods}

\begin{figure}[ptb]
\includegraphics[width=8.5 cm]{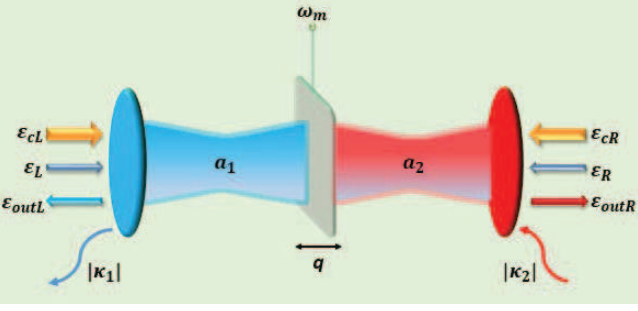}
\caption{(Color online) Schematic illustration of the optomechanical system composed of one
common mechanical resonator of frequency $\omega_{m}$ and two Fabry-P\'{e}rot cavities of
frequency $\omega_{0}$. The left cavity is passive with loss $|\kappa_{1}|$ and the right
one is active with gain $|\kappa_{2}|$. Two control fields with frequency $\omega_{c}$ and
amplitude $\varepsilon_{cL} (\varepsilon_{cR})$ together with two probe fields with frequency $\omega_{p}$
and amplitude $\varepsilon_{L} (\varepsilon_{R})$ are applied to drive the left (right) cavity.
The left (right) output field $\varepsilon_{outL} (\varepsilon_{outR})$ is of interest in this
paper.}\label{Fig:1}
\end{figure}

We consider a three-mode optomechanical system composed of a movable membrane oscillator located around
the middle position between two fixed mirrors with finite equal transmission. As shown in Fig.~\ref{Fig:1},
the movable membrane oscillator is perfectly reflected so that the system can be regarded in the form of
two Fabry-P\'{e}rot cavities with one common mechanical resonator. The two cavity modes couple to the
mechanical mode rather than couple to each other directly. Note here, the left cavity is passive while
the right one is active by adding an optical gain medium or an additional external pumping field
\cite{ptlaserhb,ptchaos,Schonleber2016NJP}. We describe the left (right) cavity mode by photon annihilation
and creation operators $a_{1}$ and $a^{\dagger}_{1}$ ($a_{2}$ and $a^{\dagger}_{2}$) with identical eigenfrequency
$\omega_{0}$. And the mechanical mode can be described by phonon annihilation (creation) operator $b (b^{\dagger})$
with eigenfrequency $\omega_{m}$. The boson operators mentioned above satisfy the commutation relation $[O_{i},O^{\dagger}_{j}]=\delta_{ij}$
($O=a_{1},a_{2},b$ and $i,j=1,2$). Two control and two probe fields are applied to drive the cavity modes
on the left and right fixed mirrors symmetrically, with the amplitude $\varepsilon_{cL}=\sqrt{2\kappa_{1}P_{1}/(\hbar\omega_{c})}$
($\varepsilon_{cR}=\sqrt{2\kappa_{2}P_{2}/(\hbar\omega_{c})}$) for the left (right) control field and
$\varepsilon_{L}=\sqrt{2\kappa_{1}\wp_{1}/(\hbar\omega_{p})}$ ($\varepsilon_{R}=\sqrt{2\kappa_{2}\wp_{2}/(\hbar\omega_{p})}$)
for the left (right) probe field, where $\kappa_{1} (\kappa_{2})$ is the decay rate of the left
passive cavity (right active cavity). In addition, we have assumed identical eigenfrequency $\omega_{c} (\omega_{p})$
for the two control (probe) fields. Then the total Hamiltonian of the system after a frame rotation with
respect to the control field frequency can be written as

\begin{eqnarray}
\mathcal{H}&=&\Delta_{0}(a^{\dagger}_{1}a_{1}+a^{\dagger}_{2}a_{2})+g_{0}(a^{\dagger}_{2}a_{2}-a^{\dagger}_{1}a_{1})(b^{\dagger}+b)\nonumber\\
&&+\omega_{m}b^{\dagger}b+i(a^{\dagger}_{1}\varepsilon_{L}e^{-i\Delta_{p} t}+a^{\dagger}_{2}\varepsilon_{R}e^{i\theta}e^{-i\Delta_{p} t}\nonumber\\
&&+\varepsilon_{cL}a^{\dagger}_{1}+\varepsilon_{cR}a^{\dagger}_{2}-h.c.)
\label{eq1}
\end{eqnarray}
where $\hbar=1$ has been set here for convenience. We introduce $\Delta_{0}=\omega_{0}-\omega_{c}$ the
detuning between cavity modes and control fields, $\Delta_{p}=\omega_{p}-\omega_{c}$
the detuning between probe fields and control fields, $\theta$ the relative phase between the
two probe fields. $g_{0}=\omega_{0}\sqrt{\hbar/(2m\omega_{m})}/L$ is the optomechanical
coupling constant due to the radiation pressure where $m$ and $L$ are the effective mass of
the mechanical resonator and the free length of the cavities, respectively.

Using the above Hamiltonian, considering relevant dissipation and quantum (thermal) noise, we can further
obtain the following Heisenberg-Langevin equations \cite{Enhancing,GenesMari,XBYan}
\begin{eqnarray}
&\dot{b}&=-i\omega_{m}b-ig_{0}(a^{\dagger}_{2}a_{2}-a^{\dagger}_{1}a_{1})-\gamma_{m}b+\sqrt{2\gamma_{m}}b_{in}\nonumber\\
&\dot{a_{1}}&=-[\kappa_{1}+i\Delta_{0}-ig_{0}(b^{\dagger}+b)]a_{1}+\varepsilon_{cL}\nonumber\\
&&\quad+\varepsilon_{L}e^{-i\Delta_{p} t}+\sqrt{2\kappa_{1}}a^{in}_{1}\nonumber\\
&\dot{a_{2}}&=-[\kappa_{2}+i\Delta_{0}+ig_{0}(b^{\dagger}+b)]a_{2}+\varepsilon_{cR}\nonumber\\
&&\quad+\varepsilon_{R}e^{i\theta}e^{-i\Delta_{p} t}+\sqrt{2\kappa_{2}}a^{in}_{2}
\label{eq2}
\end{eqnarray}
where $\gamma_{m}$ is the damping rate of the mechanical resonator which determines the mechanical
quality factor $Q=\omega_{m}/\gamma_{m}$. Moreover, $b_{in}$ is the zero-mean-value thermal noise
operator of the mechanical resonator and $a^{in}_{1} (a^{in}_{2})$ is the zero-mean-value
quantum noise operator of the left (right) cavity. Setting all the time derivatives in Eq.~(\ref{eq2})
to be zero, the steady-state values of the system variables can be obtained by neglecting the
zero-mean-value noises and the weak probe fields
\begin{eqnarray}
&b_{s}&=\frac{-ig_{0}(|a_{2s}|^{2}-|a_{1s}|^{2})}{\gamma_{m}+i\omega_{m}}\nonumber\\
&a_{1s}&=\frac{\varepsilon_{cL}}{\kappa_{1}+i\Delta_{1}}\nonumber\\
&a_{2s}&=\frac{\varepsilon_{cR}}{\kappa_{2}+i\Delta_{2}}
\label{eq3}
\end{eqnarray}
where $\Delta_{1,2}=\Delta_{0}\mp2g_{0}$Re$(b_{s})$ are the effective detunings between cavity modes
and control fields due to the feedback of the motion of the mechanical resonator. Note a relatively
weak optomechanical coupling and identical photon number in two cavities jointly lead to a negligible
term $2g_{0}$Re$(b_{s})$ compared with $\Delta_{0}$ \cite{XBYan,ptoslwl}.

To solve Eq.~(\ref{eq2}), we can safely express relevant operators as sums of the steady-state mean
values and the small quantum fluctuation terms, i.e., $O=O_{s}+\delta O$ with $O=b,a_{1},a_{2}$.
In this way, a set of linearized quantum Langevin equations can be obtained by neglecting all the
nonlinear high-order terms of fluctuation operators and the zero-mean-value quantum noise operators
after a rotating wave approximation
\begin{eqnarray}
\langle \dot{\delta b}\rangle&=&-ig_{0}(a^{*}_{2s}\langle \delta a_{2}\rangle-a^{*}_{1s}\langle \delta a_{1}\rangle)-\gamma_{m}\langle \delta b\rangle\nonumber\\
\langle \dot{\delta a_{1}}\rangle&=&-\kappa_{1}\langle \delta a_{1}\rangle+ig_{0}a_{1s}\langle \delta b\rangle+\varepsilon_{L}e^{-i\delta t}\nonumber\\
\langle \dot{\delta a_{2}}\rangle&=&-\kappa_{2}\langle \delta a_{2}\rangle-ig_{0}a_{2s}\langle \delta b\rangle+\varepsilon_{L}e^{i\theta}e^{-i\delta t}
\label{eq4}
\end{eqnarray}
where $\delta=\Delta_{p}-\omega_{m}\approx\omega_{p}-\omega_{0}$ is the detuning between the probe fields and the
cavity modes. In addition, here we have moved Eq.~(\ref{eq4}) into an interaction picture by introducing the transformation
\begin{eqnarray}
\delta b\rightarrow\delta be^{-i\omega_{m}t}, \delta b_{in}\rightarrow\delta b_{in}e^{-i\omega_{m}t}\nonumber\\
\delta a_{1}\rightarrow\delta a_{1}e^{-i\Delta_{1}t}, \delta a^{in}_{1}\rightarrow\delta a^{in}_{1}e^{-i\Delta_{1}t}\nonumber\\
\delta a_{2}\rightarrow\delta a_{2}e^{-i\Delta_{2}t}, \delta a^{in}_{2}\rightarrow\delta a^{in}_{2}e^{-i\Delta_{2}t}
\label{eq5}
\end{eqnarray}

Moreover, we assume that the three-mode system is driven by two control fields at red-detuned
sideband ($\Delta_{1}\approx\Delta_{2}\approx\omega_{m}$) and our system is studied in the resolved
sideband regime ($\omega_{m}\gg|\kappa_{1,2}|$), the common mechanical resonator possesses
a high quality factor $Q=\omega_{m}/\gamma_{m}\gg1$ and the eigenfrequency $\omega_{m}$ is assumed
to be much larger than $g_{0}|a_{1s}|$ and $g_{0}|a_{2s}|$ to ensure the validity of the rotating
wave approximation we have performed above \cite{XBYan}.

According to Eq.~(\ref{eq4}), the time-dependent oscillating terms can be removed out if the solutions are
assumed to be form: $\langle\delta O\rangle=\delta O_{+}e^{-ixt}+\delta O_{-}e^{ixt}$ where $O=b,a_{1},a_{2}$.
Note here, the component $\delta O_{+}$ possesses the same Stokes frequency $\omega_{p}$ as the probe fields
$\varepsilon_{L}$ and $\varepsilon_{R}$ and the component $\delta O_{-}$ possesses the anti-Stokes
frequency $2\omega_{c}-\omega_{p}$ via a nonlinear four wave mixing process. Then it is easy to obtain such
following results after effortless calculations
\begin{align}
&\,\delta b_{+}\,\,=\frac{-iG[n\varepsilon_{R}e^{i\theta}(k_{1}-i\delta)-\varepsilon_{L}(k_{2}-i\delta)]}{F_{1}+F_{2}}\nonumber\\
&\delta a_{1+}=\frac{G^{2}(n\varepsilon_{R}e^{i\theta}+n^{2}\varepsilon_{L})+\varepsilon_{L}(\gamma_{m}-i\delta)(k_{2}-i\delta)}{F_{1}+F_{2}}\nonumber\\
&\delta a_{2+}=\frac{G^{2}(\varepsilon_{R}e^{i\theta}+n\varepsilon_{L})+\varepsilon_{R}e^{i\theta}(\gamma_{m}-i\delta)(k_{1}-i\delta)}{F_{1}+F_{2}}
\label{eq6}
\end{align}
where $F_{1}=(\gamma_{m}-i\delta)(k_{1}-i\delta)(k_{2}-i\delta)$ and $F_{2}=G^{2}[n^{2}(k_{1}-i\delta)+(k_{2}-i\delta)]$.
We introduce $G=g_{0}c_{1s}$ the effective optomechanical coupling rate and $n=|a_{2s}/a_{1s}|^{2}$ the
photon number ratio of two cavities. Without loss of generality we set $a_{1s}$ and $a_{2s}$ real in
derivative process of Eq.~(\ref{eq6}).

According to the input-output field theory \cite{walls,scully}, we can straightforward derive the
expression of both the left and right output fields $\varepsilon_{outL}$ and $\varepsilon_{outR}$
\begin{eqnarray}
\varepsilon_{outL}&+&\varepsilon_{L}e^{-i\delta t}=2\kappa_{1}\langle\delta a_{1}\rangle\nonumber\\
\varepsilon_{outR}&+&\varepsilon_{R}e^{i\theta}e^{-i\delta t}=2\kappa_{2}\langle\delta a_{2}\rangle
\label{eq7}
\end{eqnarray}
the time-dependent oscillating terms can also be removed out similar to Eq.~(\ref{eq6}) if the solutions
here have the same form $\varepsilon_{outL}=\varepsilon_{outL+}e^{-i\delta t}+\varepsilon_{outL-}e^{i\delta t}$
and $\varepsilon_{outR}=\varepsilon_{outR+}e^{-i\delta t}+\varepsilon_{outR-}e^{i\delta t}$. Note again,
the output components $\varepsilon_{outL+}$ and $\varepsilon_{outR+}$ possess the same frequency $\omega_{p}$
as the probe fields, the output components $\varepsilon_{outL-}$ and $\varepsilon_{outR-}$ possess the
frequency $2\omega_{c}-\omega_{p}$ but not of our interest in this paper. Then according to Eq.~(\ref{eq6}) and
Eq.~(\ref{eq7}), we have
\begin{eqnarray}
\varepsilon_{outL+}&=&2\kappa_{1}\delta a_{1+}-\varepsilon_{L}\notag\\
\varepsilon_{outR+}&=&2\kappa_{2}\delta a_{2+}-\varepsilon_{R}e^{i\theta}
\label{eq8}
\end{eqnarray}

We introduce the transmission rate $T_{l}=|\varepsilon_{outR+}/\varepsilon_{L}|^{2}$
($T_{r}=|\varepsilon_{outL+}/\varepsilon_{R}|^{2}$) and the reflection rate
$R_{l}=|\varepsilon_{outL+}/\varepsilon_{L}|^{2}$ ($R_{r}=|\varepsilon_{outR+}/\varepsilon_{R}|^{2}$)
for the left (right) cavity, respectively, which are used for gauging the intensities of the output fields from
both cavities. Then we introduce $\theta_{rl}=$arg$[\varepsilon_{outL+}/\varepsilon_{L}]$
($\theta_{tl}=$arg$[\varepsilon_{outR+}/\varepsilon_{L}]$) and $\theta_{rr}=$arg$[\varepsilon_{outR+}/\varepsilon_{R}]$
($\theta_{tr}=$arg$[\varepsilon_{outL+}/\varepsilon_{R}]$) the phase of the reflection (transmission) field of the
left and right cavity, respectively, which are functions of $\omega_{p}$. In optomechanics, the group delay can be
determined by the slope of the output field phase with respect to the probe frequency, i.e., $\tau_{m}=\partial\theta_{m}/\omega_{p}$
($m=rl,tl,rr,tr$)\cite{GKH,lylpra2017}. Note a positive $\tau$ corresponding to a slow light phenomenon while a
negative $\tau$ corresponding to a fast light phenomenon on the contrary. One can see the larger the slope,
the larger the group delay, thereby the slower the light.

\begin{figure}[ptb]
\centering
\includegraphics[width=8 cm]{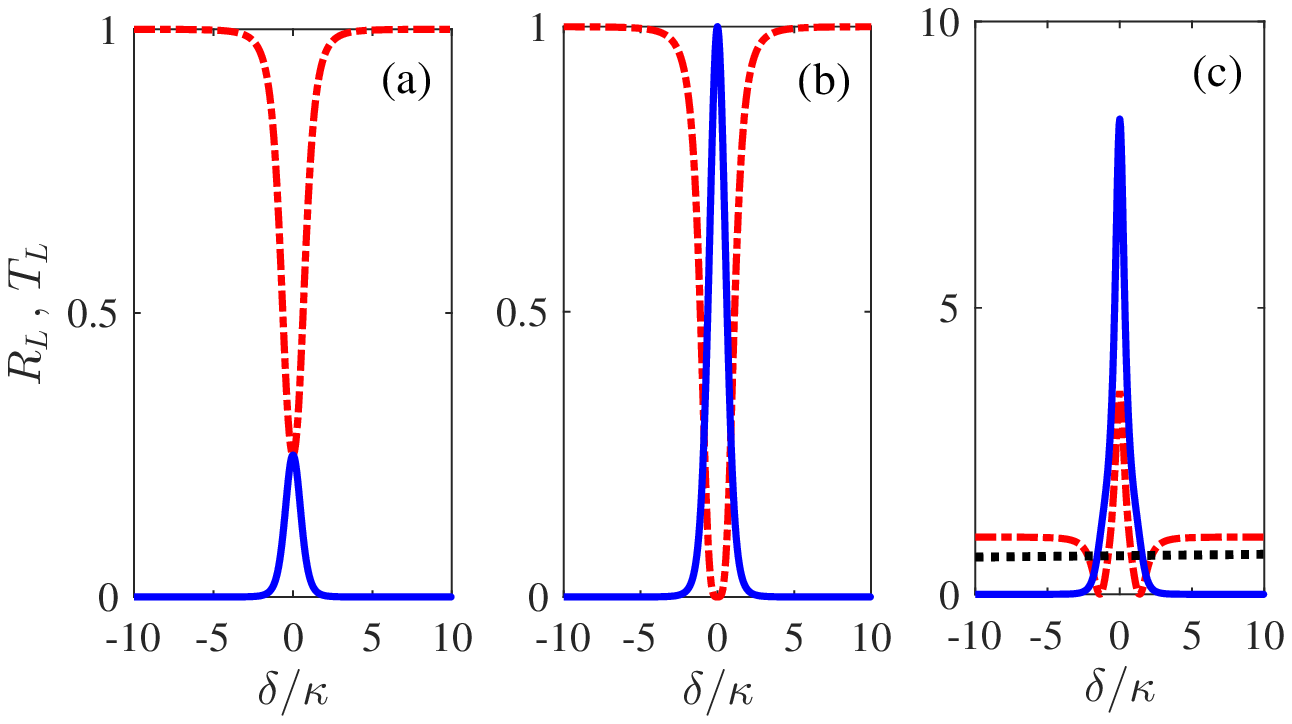}
\includegraphics[width=8 cm]{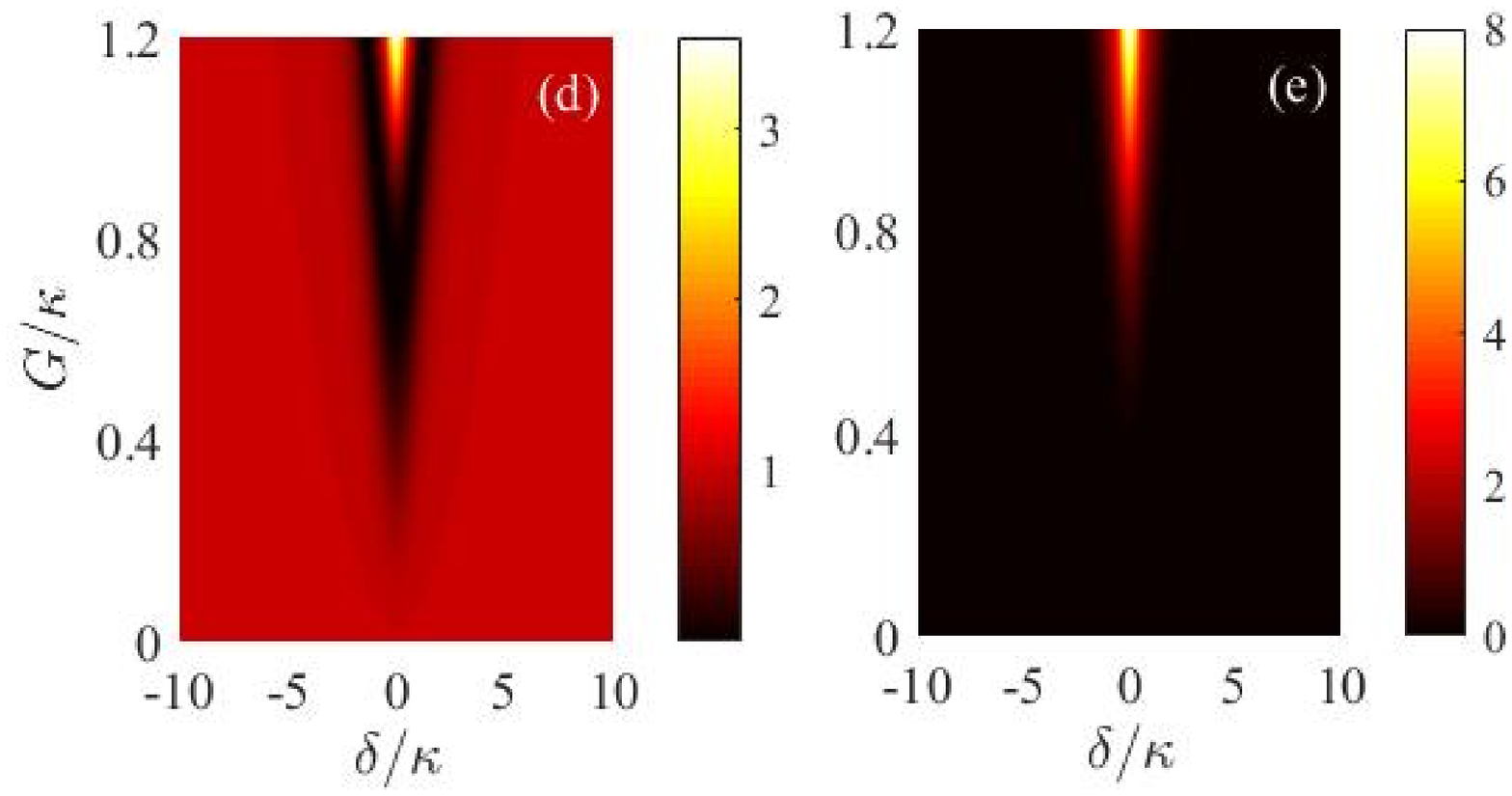}
\caption{(Color online) Reflection rate $R_{L}$ (red dot-dashed line) and transmission rate $T_{L}$
(blue solid line) versus normalized probe field detuning $\delta/\kappa$ for (a) $G=0.7\kappa$; (b) $G=\kappa/\sqrt{2}$;
(c) $G=1.2\kappa$. The black dotted line in panel (c) represents $R_{L} (T_{L})=1$ as the boundary of absorption-amplification
transition. Three-dimensional diagrams (d) and (e) corresponds to $R_{L}$ and $T_{L}$ versus $\delta$ and $G$, respectively.
Other parameters are as follows: $L=25$mm, $m=145$ng, $\kappa=2\pi\times215$kHz, $\omega_{m}=2\pi\times947$kHz and
$\lambda=2\pi c/\omega_{c}$ are the same as Ref.~\cite{XBYan}.}\label{Fig:2}
\end{figure}

\section{Results and Discussion}

Now we will study numerically how the gain effect of the active cavity impacts the photon
transport of such an optomechanical system  by setting $\kappa_{2}$ a negative value. To start
with, we should consider the assumption we mentioned above that $n=1$ is reasonable in our system.
According to Eq.~(\ref{eq3}), the photon number ratio can be indicated as
$n\sim\varepsilon^{2}_{cR}(\kappa_{1}^{2}+\omega_{m}^{2})/[\varepsilon^{2}_{cL}(\kappa_{2}^{2}+\omega_{m}^{2})]$
when $g_{0}\ll\omega_{m}$ is satisfied. Thus we can guarantee the equal photon number in the
two cavities by adjusting $\varepsilon_{cL}$ or (and) $\varepsilon_{cR}$ even if $|\kappa_{1}|\neq|\kappa_{2}|$,
and in turn, this further ensures the validity of $2g_{0}$Re$(b_{s})\ll\Delta_{0}$ in the preceding
part of the text. In this section, we find several novel and fascinating phenomena arise in different
parameter regimes which are accessible experimentally, such as (i) G-dependent photon bidirectional transport;
(ii) frequency-independent perfect reflection (FIPR); (iii) phase-dependent non-reciprocity and
(iv) fast and slow light.

\begin{figure}[ptb]
\centering
\includegraphics[width=8 cm]{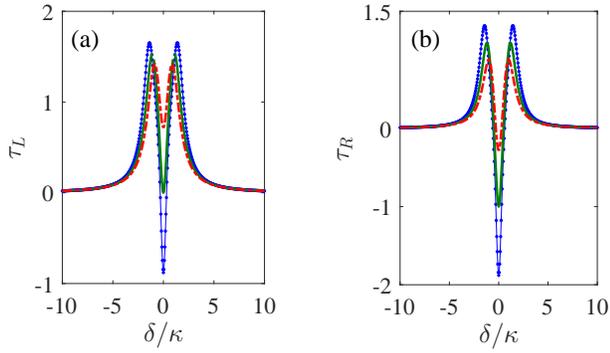}
\caption{(Color online) Group delay of the reflection field $\tau_{L}$ (a) and the transmission field $\tau_{R}$ (b)
versus normalized probe field detuning $\delta/\kappa$ for $G=0.8\kappa$ (the red dot-dashed line), $G=\kappa$
(the green solid line) and $G=1.2\kappa$ (the blue dot-solid line). All the other parameters are the same as shown in
Fig.~\ref{Fig:2} except for the parameter conditions mentioned in this section.}\label{Fig:3}
\end{figure}

\subsection{Photon bidirectional modulator}

According to Eq.~(\ref{eq6}) and Eq.~(\ref{eq8}), if no probe field is applied to the right active cavity ($\varepsilon_{cR}=0$)
and the relevant parameters of our system simultaneously satisfy $\kappa_{1}=-\kappa_{2}=\gamma_{m}=\kappa, n=1$,
the reflection and transmission rate of the left incident probe field can be written as
\begin{eqnarray}
R_{L}&=&|\frac{(\kappa+i\delta)(\kappa^{2}+\delta^{2})-2iG^{2}\delta-2\kappa G^{2}}{(\kappa-i\delta)(\kappa^{2}+\delta^{2})+2iG^{2}\delta}|^{2}\nonumber\\
T_{L}&=&|\frac{2\kappa G^{2}}{(\kappa-i\delta)(\kappa^{2}+\delta^{2})+2iG^{2}\delta}|^{2}
\label{eq9}
\end{eqnarray}
we have $R_{L}=(\kappa^{2}-2G^{2})^{2}/\kappa^{4}$ and $T_{L}=4G^{4}/\kappa^{4}$ when the
left probe field is resonant with the cavity exactly ($\delta=0$). So one can easily find that the
transmission rate is proportional to the effective optomechanical coupling rate $G$ monotonously while the
reflection rate experiences a transition from reduction to growth versus a increasing $G$.
The numerator of $R_{L}$ shows that the reflection field vanishes at $G=\kappa/\sqrt{2}$ and enters
the amplification region at the critical point $G=\kappa$. Note, by increasing $G$ from $0$ to $\kappa/\sqrt{2}$
the reflection rate decreases from $1$ to $0$ while the transmission rate increases from $0$ to $1$. Thus a
photon bidirectional modulator can be realized which enables to continuously adjust the intensity ratio between
the reflection and transmission field, means one can control the direction of the output field and the
intensity of each direction component.

\begin{figure}[ptb]
\centering
\includegraphics[width=8 cm]{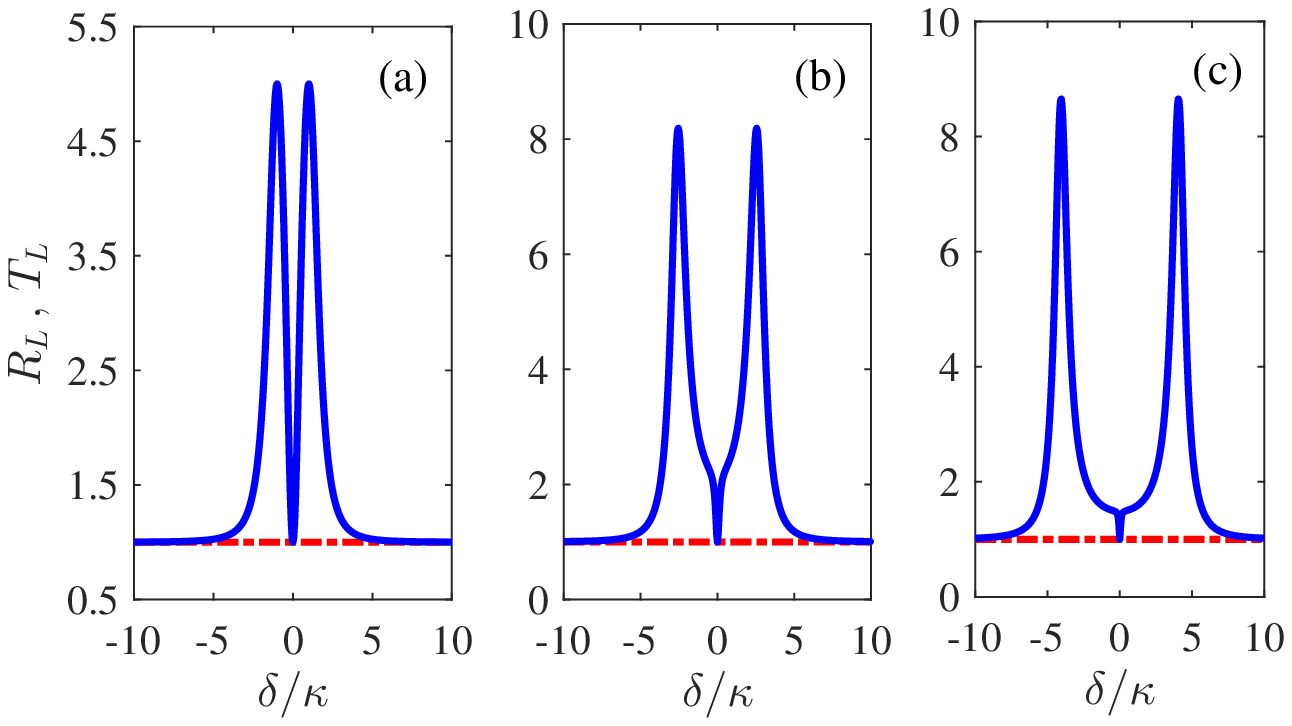}
\includegraphics[width=8 cm]{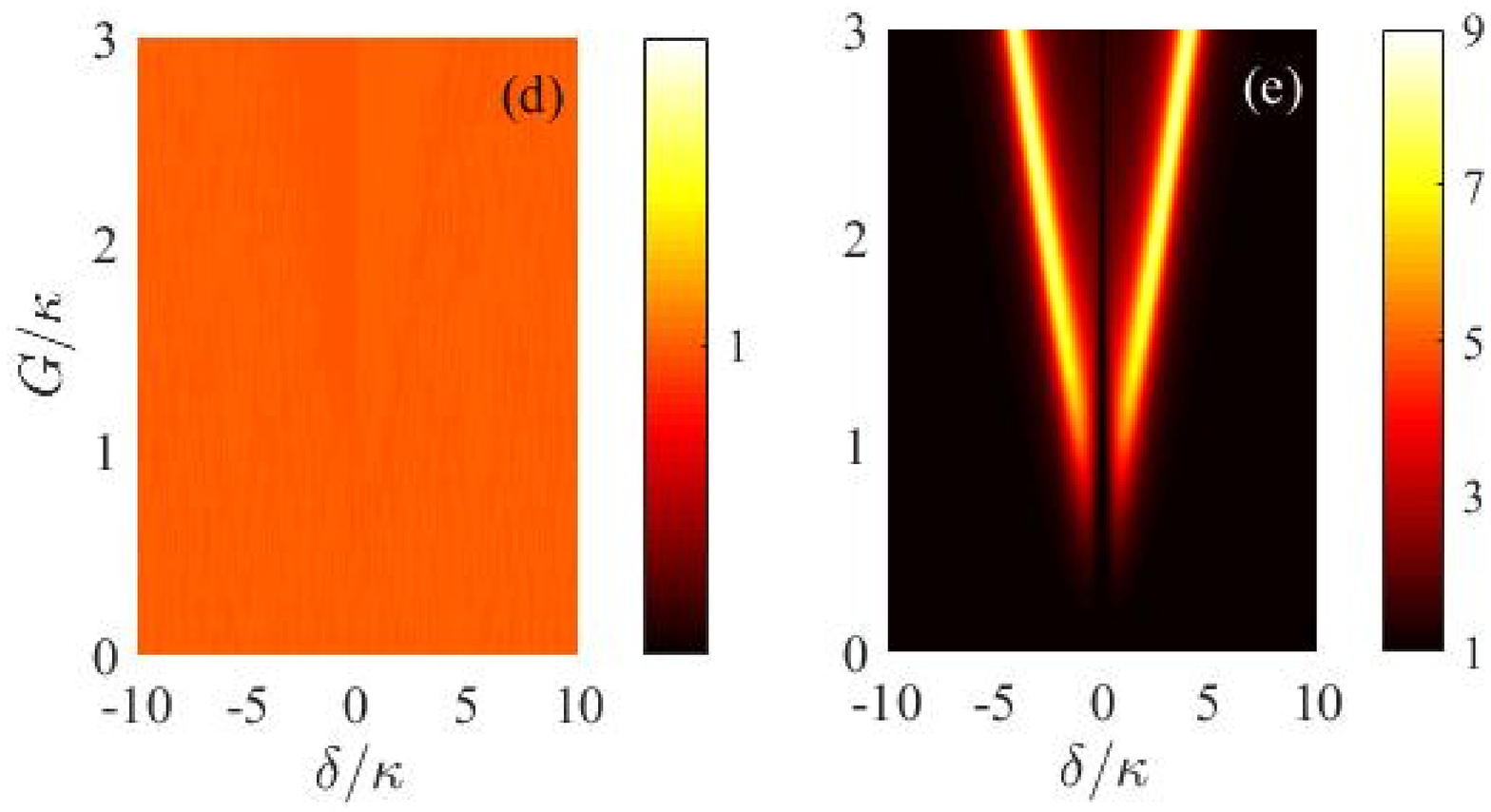}
\caption{(Color online) Reflection rate $R_{L}$ (red dot-dashed line) and transmission rate $T_{L}$
(blue solid line) versus normalized probe field detuning $\delta/\kappa$ for (a) $G=\kappa$; (b) $G=2\kappa$;
(c) $G=3\kappa$. Three-dimensional diagrams (d) and (e) corresponds to $R_{L}$ and $T_{L}$ versus $\delta$ and $G$,
respectively. All the other parameters are the same as shown in Fig.~\ref{Fig:2} except for the parameter
conditions mentioned in this section.}\label{Fig:4}
\end{figure}

Fig.~\ref{Fig:2} shows the numerical results of the photon transmission in this case. According to Fig.~\ref{Fig:2}
(a), (b) and (c), one can observe that just as we predicted analytically, the transmission rate
increases in direct proportion to $G$ with the resonance condition $\delta=0$ while the reflection rate experiences
a transition. Physically, it is because the two cavities exchange energy with the common mechanical oscillator
via radiation pressure rather than a direct coupling, there exits a competition between dissipation of the
mechanical oscillator and the indirect compensation effect. A relatively weak effective optomechanical coupling rate cannot
support a timely and effective compensation from the active cavity to the passive cavity, within this regime, a increasing
energy exchange between the left cavity mode and the mechanical mode due to the increasing $G$ leads to a greater loss
of the left passive cavity. However, a big enough $G$ amount to a fast and strong tunneling between the two cavity modes
so that an effective compensation effect arises, the indirect tunneling between the two cavity modes becomes dominant
effect, which leads to a enhanced left output field (reflection field). In addition, a coherent-perfect-transmission phenomenon proposed
in Ref.\cite{XBYan} can be observed in Fig.~\ref{Fig:2} (b) at $\delta=0$ and Fig.~\ref{Fig:2} (c) near $\delta=\pm1.3$, so besides
adjusting $G$, controllable photon transport can also be realized via adjusting the frequency of probe field.
Moreover, both the reflection and transmission field enter the amplification region while $G$ is larger than the critical point
$\kappa$ as shown in panel (c). Three-dimensional diagrams in Fig.~\ref{Fig:2} (d) and (e) exhibit the results in a more
intuitive way, from which one can clear that how to adjust the photon transport by changing the strengths of control
fields (thereby $G$) and the frequency of probe field (thereby $\delta$). In view of this, a G-dependent photon bidirectional
modulator, i.e., an all-optical photon transport switching, can be realized.

\begin{figure}[ptb]
\centering
\includegraphics[width=8 cm]{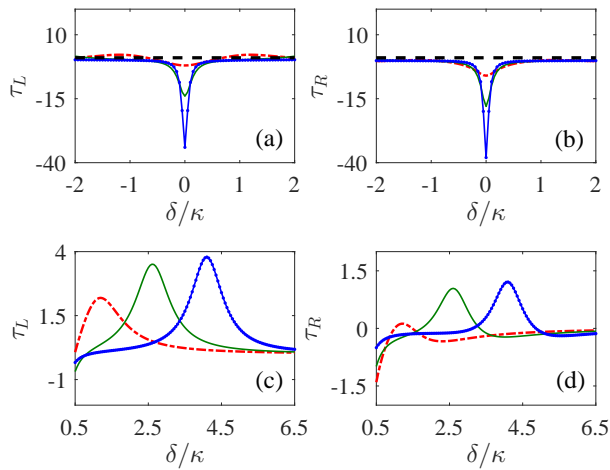}
\caption{(Color online) Group delay of the reflection field $\tau_{L}$ (a) and the transmission field $\tau_{R}$ (b)
versus normalized probe field detuning $\delta/\kappa$ for $G=\kappa$ (the red dot-dashed line), $G=2\kappa$
(the green solid line) and $G=3\kappa$ (the blue dot-solid line). The black dashed line in (a) and (b) represents
$\tau_{L} (\tau_{R})=0$ as the boundary of the fast and slow light. All the other parameters are the same as shown in
Fig.~\ref{Fig:2} except for the parameter conditions mentioned in this section.}\label{Fig:5}
\end{figure}

In this passive-active system, a tunable fast and slow light can also be realized at $\delta=0$ by adjusting the control fields.
According to the analytical expression above, once the effective optomechanical coupling rate satisfies $G>\kappa/\sqrt{2}$,
the reflection rate $R_{L}$ will be in direct proportion to $G$, so it's significative to study the fast
and slow light phenomenon of the reflection field in this region. In this case, we define $\tau_{L}=\tau_{rl}$
the reflection group delay and $\tau_{R}=\tau_{tl}$ the transmission group delay. In Fig.~\ref{Fig:3} (a), one can find a
slow-to-fast light transition of the reflection field near the critical point $G=\kappa$. The group delay $\tau_{L}$ is
inversely proportional to $G$ means a gradually faster reflected light, furthermore, $\tau_{L}$ decreases from positive
to negative and the transition point is $G=\kappa$ where $\tau_{L}=0$ as we predicted. The case in Fig.~\ref{Fig:3} (b) is slightly
different, where the transmission group delay is always negative with $G>\kappa/\sqrt{2}$, illustrating a algate fast light,
$\tau_{R}$ decreases monotonously with a increasing $G$ at $\delta=0$, so a gradually faster transmission light can also
be achieved by increase the strengths of control fields and thereby the optomechanical coupling rate $G$.

\subsection{Frequency-independent perfect reflection}

Generally, photon transport is sensitive to relevant parameters such as frequency of the probe field and
the effective optomechanical coupling. Series of novel phenomena observed in various optomechanical systems, for
instance, OMIT in $\mathcal{PT}$-symmetric optomechanical systems \cite{ptOMITlwl,ptOMITjh}, coherent perfect
absorption (CPA) in general optomechanical systems \cite{Huang,XBYan}, amplification of optical response
in $\mathcal{PT}$-symmetric-like optomechanical systems \cite{lylpra2017} and so on, arising only in the specific
parameter regions. In this section, however, we will exhibit a parameter-independent novel photon phenomenon named
frequency-independent perfect reflection (FIPR) by us.

We now consider the case that two probe fields with identical amplitudes ($\varepsilon_{L}=\varepsilon_{R}$)
incident upon the left and right cavity respectively. Based on this, when conditions $\kappa_{1}=-\kappa_{2}=\gamma_{m}=\kappa$
and $\theta=\pi, n=1$ are simultaneously satisfied. The transmission and reflection rate can be simplified as

\begin{figure}[ptb]
\centering
\includegraphics[width=7.5 cm]{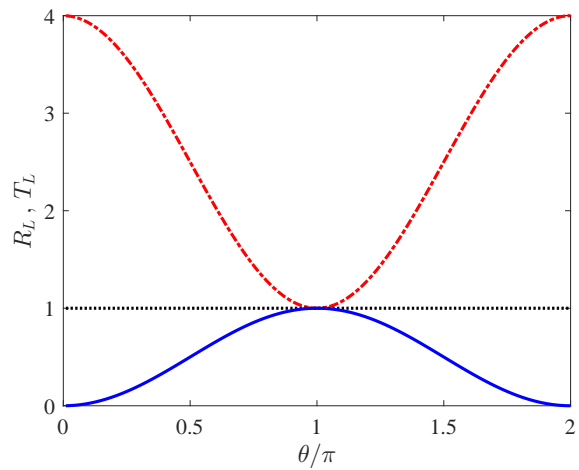}
\caption{(Color online) Reflection rate $R_{L}$ (red dot-dashed line) and transmission rate $T_{L}$
(blue solid line) versus relative phase of the probe fields $\theta$. The black dotted line represents
$R_{L} (T_{L})=1$ as the boundary of absorption-amplification transition. $G=0.5\kappa$, all the other
parameters are the same as in Fig.~\ref{Fig:5} except for $\theta$.}\label{Fig:6}
\end{figure}

\begin{eqnarray}
R_{L}&=&1\nonumber\\
T_{L}&=&|\frac{2iG^{2}\delta-(\kappa-i\delta)^{3}}{2iG^{2}\delta+(\kappa-i\delta)^{2}(\kappa+i\delta)}|^{2}
\label{eq10}
\end{eqnarray}
where we define $R_{L}=R_{l}=T_{r}$ the reflection rate and $T_{L}=T_{l}=R_{r}$ the transmission rate of the
left incident probe field in this case. So it is clear that an invariable reflection rate $R_{L}\equiv1$ is
always tenable, as shown in Fig.~\ref{Fig:4}, with no dissipation completely even if the loss rate of the cavity
and mechanical oscillator is big enough, illustrating a perfect reflection in a fairly broad range of the probe
frequency which has a promising prospect for photonic device fabrication. In addition, a remarkable amplification
of the transmission field can be observed in Fig.~\ref{Fig:4}. According to panels (a), (b) and (c), one can find
that the degree of amplification is in direct proportion to $G$ and a controllable amplification can be realized.
Panels (d) and (e) provide a more intuitive manifestation of our results, shows that the reflection field is also
irrelevant to $G$ according with Eq.~(\ref{eq10}) and the amplified transmission field can be controlled by both the strengths
of control fields and the frequencies of probe fields.

The case in Fig.~\ref{Fig:5} is much more different compared with that in Fig.~\ref{Fig:3}. Here we define $\tau_{L}=\tau_{rl}=\tau_{tr}$
the reflection group delay and $\tau_{R}=\tau_{tl}=\tau_{rr}$ the transmission group delay of the left incident probe field.
In panels (a) and (b), similar to Fig.~\ref{Fig:3}, both $\tau_{L}$ and $\tau_{R}$ decrease with a increasing $G$ at $\delta=0$,
where $R_{L}=T_{L}=1$, illustrating a tunable fast light with lossless reflection and transmission field can be realized in
this system when the probe fields are resonant with cavity modes. Moreover, in panels (c) and (d), both the reflection and
transmission group delay are in direct proportion to a increasing $G$ near the optimal value of $\delta$ where the transmission
field has the maximal amplification effect according to Fig.~\ref{Fig:4} ($\delta\sim1, \delta\sim2.5$ and $\delta\sim4$
corresponding to $G=\kappa, G=2\kappa$ and $G=3\kappa$, respectively). So we can realize a tunable slow light accompanied by a
amplified transmission field and a lossless reflection field. In a word, by adjusting the strengths of control fields and the
frequencies of probe fields, one can realize both fast and slow light for both the reflection and transmission fields.

Up to now in this section, we have chosen a fixed relative phase of the probe fields $\theta=\pi$. Then we will show that
except for adjusting $G$ and $\delta$, $\theta$ is also an important control parameter for realizing a all-optical
switching. If the condition $\varepsilon_{L}=\varepsilon_{R}, \kappa_{1}=-\kappa_{2}=\gamma_{m}=\kappa, n=1$ is met,
then we have
\begin{eqnarray}
R_{L}&=&\big|1-\frac{2G^{2}(1+e^{i\theta})}{\kappa^{2}}\big|^{2}\nonumber\\
T_{L}&=&\big|e^{i\theta}+\frac{2G^{2}(1+e^{i\theta})}{\kappa^{2}}\big|^{2}
\label{eq11}
\end{eqnarray}
with resonant probe fields ($\delta=0$). According to Eq.~(\ref{eq11}), $R_{L}=0, T_{L}=|1+4G^{2}/\kappa^{2}|^{2}$ when $\theta=2m\pi (m=0,1,2,...)$
and $R_{L}=T_{L}=1$ when $\theta=m'\pi (m'=1,3,5,...)$, as shown in Fig.~\ref{Fig:6}. Moreover, both the reflection and transmission rate
can be continuously adjusted via $\theta$ and the amplification effect of the transmission field relies on $G$. Physically, it is
because by adjusting the relative phase $\theta$, the interference effect between the two probe fields can be changed, and this also
impacts the compensation effect between the two cavity modes. In view of this, a phase-dependent all-optical switching can be realized
by adjusting the relative phase of the probe fields and the strengths of the control fields.

\section{Conclusions}

In summary, we have studied theoretically how to control photon transport in a passive-active optomechanical
system driven by two strong control fields and two weak probe fields. According to the main text,
in virtue of the gain effect of the active cavity, series of novel and valuable phenomena arise, illustrating the
realization of an all-optical photon transport switching. By adjusting the strengths of control fields, the frequencies
of probe fields and the gain rate of the active cavity, we can realize: (i) a photon bidirectional modulator relying
on the effective optomechanical coupling rate $G$; (ii) frequency-independent perfect reflection (FIPR) which subverts tradition
so that the complete reflection in whole frequency region becomes possible. (iii) a phase-dependent photon switching by
adjusting the relative phase of the probe fields. Furthermore, tunable fast and slow light can be realized in this passive-active
optomechanical system by accurately adjusting the input fields and the gain rate of the active cavity in a
experimentally reasonable region.

\section*{ACKNOWLEDGMENTS}

Thanks Dr. Chu-Hui Fan for helpful discussions and Dr. Hong-Zhi Shen for constructive suggestions. Jin-Hui Wu is supported by
National Natural Science Foundation of China under Grants No. 61378094, 11534002 and 11674049). Yi-Mou Liu is supported by
National Natural Science Foundation of China under Grant No. 11704063. Lei Du and Yan Zhang acknowledge support from
National Natural Science Foundation of China under Grant No. 11704064.

\end{document}